\documentclass[useAMS,usenatbib]{mn2e}

\usepackage{psfig}

\title[Optical spectroscopy of GX\,9$+$9]{Optical spectroscopy of the low mass X-ray binary GX\,9$+$9}
\author[R. Cornelisse et~al.]{R. Cornelisse$^{1,2}$\thanks{E-mail:
corneli@iac.es}, D. Steeghs$^{3,4}$, J. Casares$^{1}$, P.A. Charles$^{5,2}$,  A.D. Barnes$^{2}$,
\newauthor
R.I. Hynes$^{6}$, K. O'Brien$^{7}$\\
$^{1}$Instituto de Astrofisica de Canarias, Via Lactea, La Laguna E-38200, Santa Cruz de Tenerife, Spain\\
$^{2}$School of Physics and Astronomy, University of Southampton, Highfield, Southampton SO17 1BJ, UK\\
$^{3}$Harvard-Smithsonian Center for Astrophysics, 60 Garden Street, Cambridge, MA 02138, USA\\
$^{4}$Department of Physics, University of Warwick, Coventry, CV4 7AL, UK\\
$^{5}$ South Africa Astronomical Observatory, P.O.Box 9.Observatory 7935, South Africa\\
$^{6}$Department of Physics and Astronomy, 202 Nicholson Hall, Louisiana State University, Baton Rouge, LA 70803, USA\\
$^{7}$European Southern Observatory, Casilla 19001, Santiago 19, Chile\\
}

\begin{document}

\date{Accepted  Received ; in original form }

\pagerange{\pageref{firstpage}--\pageref{lastpage}} \pubyear{2004}

\maketitle

\label{firstpage}

\begin{abstract}
  Phase-resolved medium resolution VLT spectroscopy of the low mass
  X-ray binary GX\,9$+$9 has revealed narrow C\,III emission lines
  that move in phase relative to our new estimate of the ephemeris,
  and show a velocity amplitude of 230$\pm$35 km s$^{-1}$. We identify
  the origin of these lines as coming from the surface of the donor
  star, thereby providing the first estimate of the mass function of
  $f(M_1)$$\ge$0.22$M_\odot$. Rotational broadening estimates together with
  assumptions for the mass donor give 0.07$\le$$q$$\le$0.35 and
  182$\le$$K_2$$\le$406 km s$^{-1}$. Despite a low mass ratio, there
  is no evidence for a superhump in our dataset.
  Doppler maps of GX\,9$+$9 show the presence of a stream overflow,
  either in the form of material flowing downward along the accretion
  disk rim or in a similar fashion as occurs in high mass transfer
  rate cataclysmic variables known as the SW\,Sex stars. Finally we
  note that the Bowen region in GX\,9$+$9 is dominated by C\,III
  instead of N\,III emission as has been the case for most other X-ray
  binaries.
\end{abstract}


\section{Introduction}

One of the main aims of optical observations of low mass X-ray
binaries (LMXBs) has always been to find a signature of the donor
star, and thereby constrain the masses of both components.  However,
until recently it were mainly the LMXBs that show transient outbursts
for which it was possible to determine the mass function (see e.g.
Charles \& Coe 2006). During a transient outburst the optical emission
is completely dominated by the X-ray irradiated accretion disk, and it
is only when the source has returned to quiescence (or at least the X-ray 
flux has decreased by several orders) that the donor star
becomes directly visible. That is why, until recently, there were no
strong constraints on the mass of the donor and compact object for
most LMXBs that are persistently bright, although optical counterparts
have been known for many systems for over 20 years.

This situation changed when Steeghs \& Casares (2002) developed a new
technique to detect the companion star in persistent LMXBs. Using
phase-resolved blue spectroscopy they revealed narrow emission line
components that were interpreted as coming from the irradiated
companion star of Sco\,X-1. This enabled them for the first time to
determine the mass function. These narrow features were most obvious
in the Bowen blend (a blend of N\,III 4634/4640 \AA\ and C\,III
4647/4650 \AA\ lines), which is the result of UV fluorescence due to
the hot inner disk for the N\,III lines or photo-ionization and
subsequent recombination for the C\,III lines (McClintock et~al.
1975). This led to the discovery of such signatures in other
persistent LMXBs such as X\,1822$-$371, 4U\,1636$-$53 and
4U\,1735$-$44 (Casares et~al. 2003, 2006), or transient sources where
the system parameters could not be constrained in quiescence such as
GX\,339$-$4 or Aql\,X-1 (Hynes et~al. 2003, Cornelisse et~al. 2007).

GX\,9$+$9 is a persistent LMXB that is bright in both optical as well as 
X-rays and is another good candidate for a Bowen study. Its compact
object is thought to be a neutron star due to its X-ray spectral and
timing properties (e.g. Hasinger \& van der Klis 1989), although direct
evidence such as Type\,I X-ray bursts or pulsations have not been
observed thus far. The optical counterpart of GX\,9$+$9 was
identified by Davidsen et~al. (1976) as a $V$$=$16.6 blue object. A
4.2 hour periodic modulation in X-rays was detected that was
interpreted as the orbital modulation (Hertz \& Wood 1988), and
confirmed in the optical by Schaefer (1990). From its aperiodic
variability at very low frequencies, Reig et~al. (2003) speculated
that GX\,9$+$9 has an early type (earlier than G5) companion. Haswell
\& Abbott (1994) suggested that GX\,9$+$9 is a persistent superhumper,
based on its supposed mass ratio ($q$$<$0.28), the optical
modulations, and the change in the shape of the optical
lightcurves. Superhumps are periodic optical modulations first
observed in SU\,UMa dwarf novae (see Warner 1995), but also in black
hole soft X-ray transients (O'Donoghue \& Charles 1996), and are due
to an eccentric accretion disk that is undergoing precession (e.g.
Whitehurst 1988).

Recent simultaneous X-ray and optical observations by Kong
et~al.(2006) did not detect a significant X-ray/optical correlation.
Although they could not confirm the X-ray modulation at the orbital
period, it is present in the RXTE/ASM data (Levine
et~al. 2006a,b). Time resolved X-ray spectroscopy by Kong
et~al. (2006) showed that a two-component spectral model was needed to
fit the X-ray emission. The best model suggested that an optically
thick boundary layer which is partly obscured (by either the inner
disk or thickened disk) and an extended hot corona are present in
GX\,9$+$9.

\begin{figure}\begin{center}
\psfig{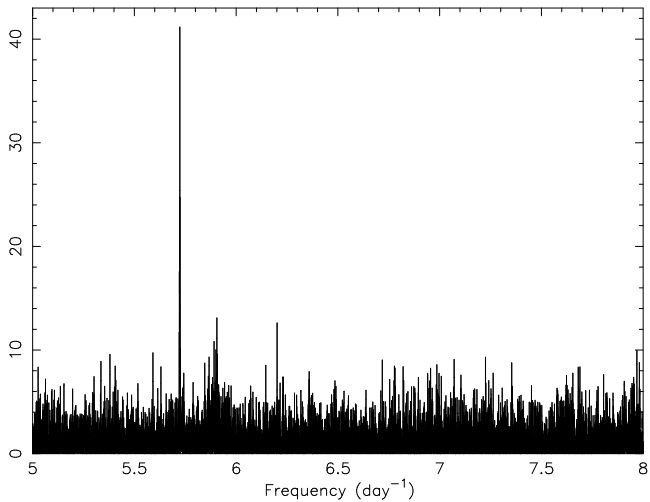}
\end{center}
\caption{Scargle normalised power spectrum of the RXTE/ASM data of 
GX\,9$+$9, showing a strong peak at the expected (orbital) period of 
$\simeq$4.2 hrs.
\label{period}}
\end{figure}

In this paper we present VLT medium resolution blue spectroscopy of
GX\,9$+$9, and we derive a new ephemeris based on X-ray data. We
determine radial velocity curves and Doppler maps of the most
important lines and use these to derive the first constraints on the
system parameters.

\section{observations and data reduction}

\begin{figure}\begin{center}
\psfig{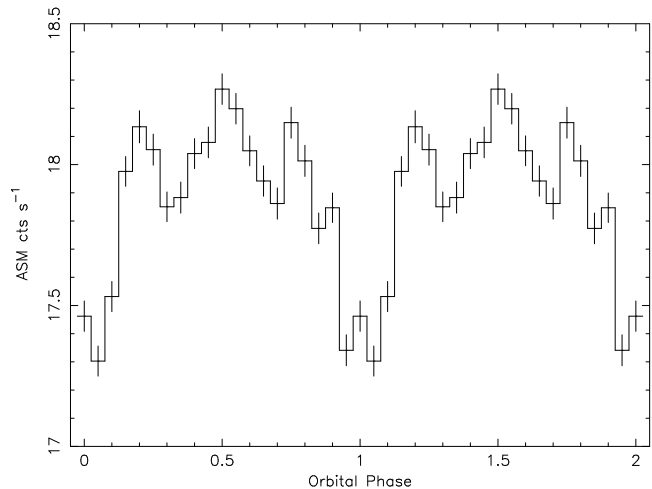}
\end{center}
\caption{Folded RXTE/ASM X-ray lightcurve of GX\,9$+$9 on the orbital
period, where phase zero is defined as the minimum in the lightcurve. 
For clarity, the lightcurve is shown twice.
\label{folded}}
\end{figure}

On June 24 2003 we obtained two spectra of GX\,9$+$9 in order to
establish the strength and velocity structure of its Bowen blend. This
was followed by a phase-resolved study from 25 to 27 May 2004 using
the FORS\,2 spectrograph attached to the VLT Unit 4 (Yepun Telescope)
at Paranal Observatory (ESO). During each night in 2004 we observed
GX\,9$+$9 for approximately one full orbit with the 1400V
volume-phased holographic grism, resulting in a total of 58 spectra
(+2 in 2003) with an integration time of 600\,s each. We used a slit
width of 0.7$''$, giving a wavelength coverage of
$\lambda$$\lambda$4514-5815 with a resolution of 70 km s$^{-1}$
(FWHM). We also observed GX\,9$+$9 once with a wide, 2.0 arcsec slit
width. The seeing during these three nights varied between 0.5 and 2.1
arcsec, and was 1.1 arcsec during the observation with the wide
  slit. The slit was orientated at a position angle of
-58.5$^{\circ}$ to include a comparison star in order to correct for
slit losses. Since we do not have a flux standard, we were not able to
correct for the instrumental sensitivity curve. During the daytime He,
Ne, Hg and Cd arc lamp exposures were taken for wavelength
calibration. We determined the pixel-to-wavelength scale using a 4th
order polynomial fit to 20 reference lines giving a dispersion of 0.64
\AA\,pixel$^{-1}$ and rms scatter $<$0.05 \AA. We de-biased and
flat-fielded all the images and used optimal extraction techniques to
maximise the signal-to-noise ratio of the extracted spectra (Horne
1986), using the routines from the PAMELA package. We also
corrected for any velocity drifts due to instrumental flexure (always
$<$5 km s$^{-1}$) by cross-correlating with the sky spectra. The
  subsequent analysis was performed using the package MOLLY.

\section{Data analysis}

\subsection{Orbital period}

\begin{figure*}\begin{center}
\psfig{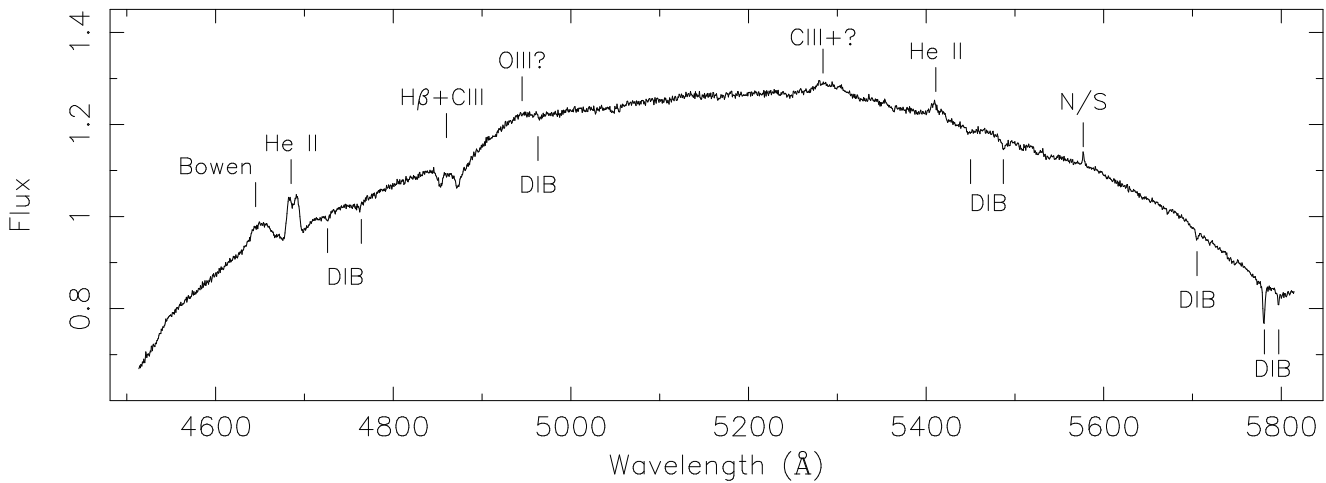}
\end{center}
\caption{Average normalised spectrum of GX\,9$+$9 from our 3 nights of
VLT data. We have indicated the most prominent emission lines and
interstellar diffuse bands (DIB). Furthermore we have indicated a sky
line (N/S) that has not been properly removed.
\label{spectrum}}
\end{figure*}

The first, and only, optical orbital ephemeris of GX\,9$+$9 was
published by Schaefer (1990) over 16 years ago. Unfortunately, due to
the propagation of the phase uncertainty, this ephemeris cannot be
projected to the present. However, Levine et~al. (2006a,b) showed that
it is possible to detect the orbital period in the publicly available
X-ray lightcurve provided by the All Sky Monitor onboard the RXTE
satellite (RXTE/ASM).  Therefore, we decided to derive a new ephemeris
from this dataset by also searching for the 4.2 hrs period and
determine a new zero point, using the individual dwell, full 2-12
keV band-pass observations. We applied a barycentre correction to
the data using the IDL routine HELIO$\_$JD and searched for the
orbital period with the Lomb-Scargle technique (Scargle 1982). Only
one significant peak is observed, at 4.19345(2) hrs (see
Fig.\,\ref{period}). This value is similar to that of Levine et~al.
(2006b), but note that this period is slightly, but according to the
formal errors significantly (4.7$\sigma$), shorter than the
photometric period of 4.1958$\pm$0.0005 hrs detected by Kong et~al.
(2006). In order to see if this difference is due to a change in
period, we have also split the RXTE/ASM dataset in two equal lengths
and determined the orbital period for each. However, both periods are
equal, and we conclude that the orbital period has not significantly
changed over the $\simeq$11 yr baseline of the RXTE/ASM data.

We phase-folded all the data on the observed period and fitted a sine
curve, using a date close to our observations as the zero point. We
then determined the minimum in our folded RXTE/ASM lightcurve to be JD
2453151.464(3), and define this as orbital phase zero. In
Fig.\,\ref{folded} we show the folded lightcurve.  Although the
lightcurve exhibits a modulation of $\simeq$4\%, comparable to that
observed by Hertz \& Wood (1988), Levine et~al.(2006a) reported that
the amplitude changes over time and even increased up to 18\% after
Jan 2005.  Note that we define our orbital phase to be half an orbit
removed from that used by Schaefer (1990), i.e. more in line with the
standard definition where phase zero corresponds to inferior
conjunction of the secondary star.  This gives a new ephemeris (in HJD) of:\\

$T$=2453151.464(3)+0.1747272(8)$E$\\

\noindent which we will use in the remainder of this paper. In
Sects.\,3.3 and 3.5 we will determine the ephemeris in three
different, independent, ways. They are all consistent with the above
ephemeris, giving us confidence that it represents a reliable orbital
phasing. However, we must always be careful with the assumption that
the minimum in the lightcurve corresponds to inferior conjunction of
the secondary star.

\subsection{Flux calibration and spectrum}

In order to correct for slit losses due to the variable seeing we used
a comparison star placed on the slit. We divided all comparison star
spectra by its wide-slit spectrum, and fitted them with a low-order
spline. The spectra of GX\,9$+$9 were subsequently divided by the
comparison star splines (corresponding to the same observation) in
order to get the final spectra corrected for slit losses. Note that since 
we do not have a flux standard all our spectra have relative fluxes.

Figure\,\ref{spectrum} shows the average spectrum of GX\,9$+$9
normalised to unity. It is dominated by high excitation emission lines
from HeII $\lambda$4686, $\lambda$5411 and Bowen
$\lambda$$\lambda$4630-4650. A strong absorption feature with emission
structure superposed is present around H$\beta$. Several absorption
lines can also be observed that are mainly due to interstellar bands
(indicated in Fig.\,\ref{spectrum} as DIB). A close inspection of the
Bowen region shows that there is little emission below 4640 \AA, i.e.
the part of the blend where all the dominant N\,III fluorescence lines
should be. In order to quantify this absence of N\,III we measured the
centroid of the Bowen region using a Gaussian, and find a central
wavelength of 4648.0$\pm$0.2 \AA\, (while He\,II has a central
wavelength of 4686.7$\pm$ 0.2 \AA). This central wavelength is much
higher than was the case for e.g. V801\,Ara (4643.2 \AA) and V926\,Sco
(4641.5 \AA; Casares et~al. 2006).  Although there might be some
N\,III present, it does suggest that the Bowen blend is dominated by
C\,III in GX\,9$+$9. We therefore checked where other strong C\,III
transitions should occur in our spectra. There should be one emission
line at 4860 \AA, with a transition strength comparable to those in
the Bowen region. Although this is close to H$\beta$ we will show in
Sect.\,3.5 that we have most likely detected it, although the presence
of He\,II $\lambda$4859 can also not be ruled out.  Another strong
C\,III transition occurs at 5305 \AA, and is also present in our
spectrum (indicated in Fig.\,\ref{spectrum}), although the width of
the line suggests that it is a blend with other unknown lines.
Finally, we checked if any strong N\,III or He\,I lines are present in
our spectral range (apart from the N\,III lines in the Bowen region).
Although there should be 3 strong N\,III lines at 4867, 5059 and 5320
\AA\, and 2 He\,I lines at 4922 and 5015 \AA, none of these appear to
be present.

\subsection{Optical lightcurves}

\begin{figure}\begin{center}
\psfig{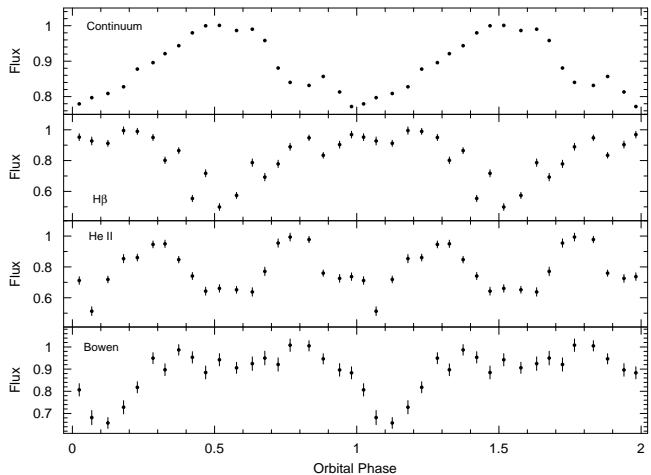}
\end{center}
\caption{Average lightcurves of GX\,9$+$9 with the maximum normalised
to unity (also for H$\beta$, despite negative flux levels) and folded
on the orbital period of 4.19 hrs for the continuum, H$\beta$, He\,II
$\lambda$4686 and the Bowen blend respectively. For clarity each
lightcurve is plotted twice.
\label{light}}
\end{figure}

We used the orbital ephemeris derived above (Sect. 3.1) to compute the
orbital phase of each spectrum. We integrated the continuum flux from
each individual spectrum, phase-folded them into 20 bins, and
normalised the peak flux to unity in order to create the lightcurve
that is shown at the top of Fig.\,\ref{light}. The first thing to note
is that the minimum in the optical continuum lightcurve corresponds
very well with orbital phase 0 as defined by the X-ray modulation, and
thereby strengthening our interpretation that the X-ray ephemeris
represents inferior conjunction of the secondary. Furthermore, we also
note that the flux varies by $\simeq$20\%, comparable to previous
observations (e.g. Schaefer 1990, Kong et~al. 2006).

We have also subtracted the continuum lightcurve from each spectrum in
order to derive light curves from the main emission or absorption
lines, and these are also shown in Fig.\,\ref{light}. We do note that
on average the flux in the lines vary by 40\%, i.e. the variation is
twice as strong as in the continuum. The flux of the H$\beta$
absorption line shows more or less a sinusoidal variation, with
minimum (i.e. maximum absorption) occurring around orbital phase 0.5
(in anti-phase with the continuum emission). The variation of
He\,II\,$\lambda$4686 and the Bowen blend is more complex. He\,II
$\lambda$4686 appears to be a combination of the H$\beta$ and
continuum lightcurves, i.e. showing a minimum around orbital phase 0.5
and a second minimum at phase 1.0. The Bowen emission, on the other
hand, looks similar to the continuum emission, although its single
minimum is at a slightly later orbital phase of $\simeq$1.1.

\subsection{Trailed spectra and radial velocity curves}

\begin{figure}
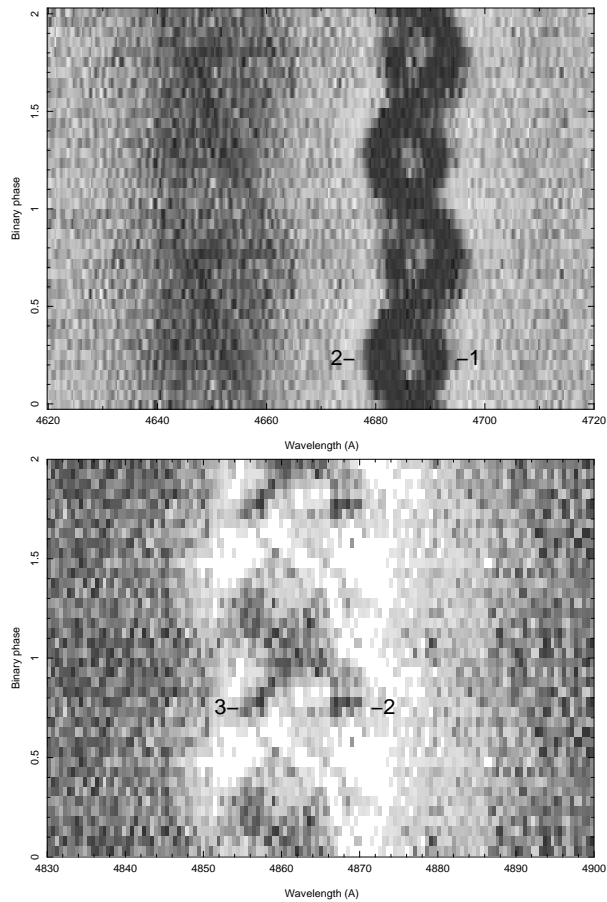
\begin{center}
\psfig{figure=trail.2.ps,angle=-90,width=8.0cm}
\psfig{figure=trail_hbeta.ps,angle=-90,width=8.0cm}
\caption{Trail of the Bowen blend and He\,II $\lambda4686$ line (top
panel), and H$\beta$ (bottom panel), folded on the orbital period
and plotted twice for clarity. The numbers indicated in the 
trails correspond to the component numbers in Table\,\ref{sine}. The 
light areas in the H$\beta$ trail are the absorption feature, while 
dark regions are the emission features.
\label{trail}}
\end{center}\end{figure}

We have phase-folded all data into 20 bins, and plotted a trail of the
region around He\,II\,$\lambda$4686 and the Bowen blend in
Fig.\,\ref{trail} (top panel). The He\,II\,$\lambda$4686 line clearly
consists of two separate components that we have indicated with 1
  and 2 in Fig.\,\ref{trail} and correspond to the component number
  in Table\,\ref{sine}. Both components show a sine-like modulation in
anti-phase with each other. Note that these oppositely phased S-waves
also produce double peaks, but these are completely different from the
more conventional disk dominated line profiles where the peaks move
together in phase. There is no strong broad component visible,
indicating that most of the He\,II $\lambda$4686 observed is coming
from two distinct regions in the binary.  We estimated the centre of
each emission component as a function of orbital phase using two
Gaussians, and then fitted each component with a sine curve, obtaining
the parameters given in Table\,\ref{sine}. Fig.\,\ref{trail} already
suggests that component 2 has twice the velocity amplitude of
component 1, and this is confirmed in Table\,\ref{sine}.

\begin{figure}\begin{center}
\psfig{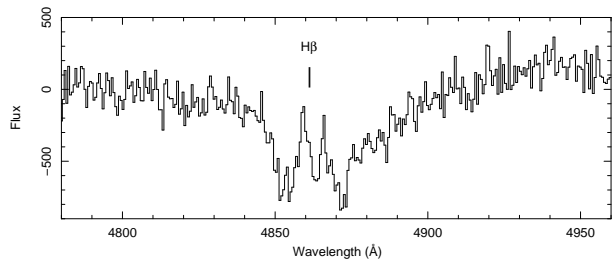}
\caption{Average spectrum of the H$\beta$ region of GX\,9$+$9 around
orbital phase 0.5, clearly showing that the main emission components
superposed on the broad absorption are still present.
\label{abs}}
\end{center}\end{figure}

We also examined the region around H$\beta$ in more detail, and show a
trail in Fig.\,\ref{trail} (bottom panel). Due to the presence of the
absorption feature this region is more complex. However, we do notice
that the emission feature also appears to consist of two components
(indicated with 2 and 3), more or less in anti-phase with each other.
These components are clearly visible in the individual spectra (see
e.g. Fig.\,\ref{abs}), and made us confident that we could again fit
two Gaussians to the emission components (see also Table\,\ref{sine}
for the best fitting sine curves, which we have labelled C\,III and
H$\beta$). We note that the strongest feature (called component 2)
appears to originate from the same region as component 2 of He\,II
$\lambda$4686, and we will discuss this in more detail in Sect.\,3.5.
On the other hand, if we assume that component 3 is also H$\beta$
emission, its average velocity, $\gamma$, is completely different
compared to all the other components listed in Table\,\ref{sine}
($\gamma$ would be -16 km s$^{-1}$ compared to $\simeq$95 km
s$^{-1}$). However, if we assume this line is C\,III $\lambda$4859.65
and/or He\,II $\lambda$4859.3 (see Sect.\,3.2) $\gamma$ becomes
comparable with that derived for the other emission lines. We,
therefore, conclude that component 3 is most likely tracing a
combination of C\,III and He\,II.  In Sect.\,3.5 we will show that
this region is most likely dominated by C\,III and H$\beta$ emission,
and despite its complicated appearance this trail is very similar to
He\,II $\lambda$4686.

In the H$\beta$ trail (Fig.\,\ref{trail}) it appears that no net
emission components are present at orbital phase 0.5. However, given
that the absorption feature hardly moves with orbital phase, we expect
the emission feature to be at the deepest point of the absorption
feature at orbital phase 0.5. If the peak flux of the emission feature
is below the continuum level it would be difficult to detect its
signature in the trail. In Fig.\,\ref{abs} we show the average
spectrum around orbital phase 0.5, and it is clear that the H$\beta$
and C\,III/He\,II emission components are still present (but have peak
fluxes below the continuum level).

\begin{table}\begin{center}
\caption{Kinematics of GX\,9$+$9 emission line structures obtained by
best fitting sine-curves to the double peaked structure observed in
He\,II $\lambda$4686 and H$\beta$ regions. The component numbers for
He\,II correspond to the numbers indicated in the He\,II trail of 
Fig.\,\ref{trail}, and H$\beta$ component 2 has the same origin as 
component 2 in He\,II.
\label{sine}}
\begin{tabular}{ccccc}
\hline
& component         & $K$         & $\gamma$ & $\phi$ \\
&    \#     & (km s$^{-1}$) & (km s$^{-1}$) &\\
\hline
He\,II   & 1& 217$\pm$5 & 97$\pm$5   & 0.07$\pm$0.05\\
He\,II   & 2& 402$\pm$3 & 91$\pm$3   & 0.55$\pm$0.05\\
C\,III   & 3& 260$\pm$5 & 87$\pm$5   & 0.02$\pm$0.05\\
H$\beta$ & 2& 398$\pm$8 & 100$\pm$6  & 0.49$\pm$0.05\\
\hline
\multicolumn{5}{l}{$K$=semi-amplitude,$\gamma$= off-set to rest-wavelength,}\\
\multicolumn{5}{l}{$\phi$=phase zero}\\
\end{tabular}
\end{center}\end{table}

We also created trails of each individual night by folding them on the
ephemeris determined in Sect.\,3.1. If GX\,9$+$9 is a superhumper, it
would most likely show a similar spectroscopic signature as other
superhumping LMXBs, namely a movement in the nightly mean emission
line profiles (see e.g. Zurita et~al. 2002). Therefore, we did test if
movement is present, but no shift in the line centroid was detected
between the different nights. Also a comparison between the 2003 and
2004 spectra shows no clear shift that would unambiguously point
toward a spectroscopic signature of a superhump in GX\,9$+$9.

\begin{figure}\begin{center}
\psfig{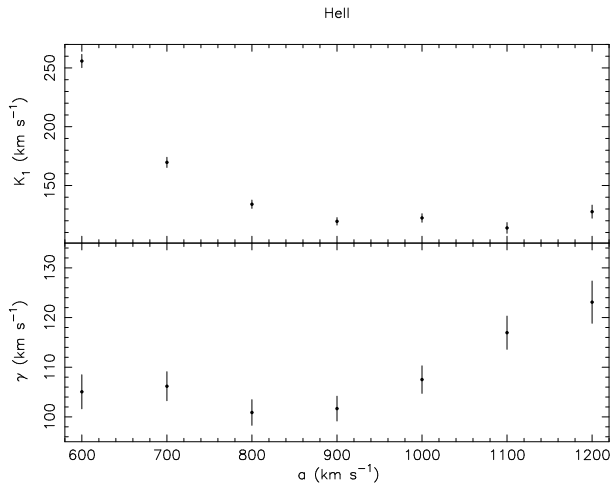}
\caption{Diagnostic diagram for GX\,9$+$9 of He\,II $\lambda$4686 to
follow the estimates of the radial velocity of the compact object,
$K_1$, and the systemic velocity, $\gamma$, as a function of Gaussian
separation, a.
\label{wings}}
\end{center}\end{figure}

In order to determine the radial velocity of the compact object one
can apply the double-Gaussian technique (Schneider \& Young 1980). We
must note that this technique only works for a standard double peaked
emission profile disk, which is clearly not observed in GX\,9$+$9 (see
Fig.\,\ref{trail}). We still decided to apply the technique to He\,II
$\lambda$4686 using velocities that are as far as possible from the
prominent emission features in the trail, in the hope that their
contribution is negligible. We used a double Gaussian bandpass with
FWHM of 100 km s$^{-1}$ and Gaussian separations between 600 and 1200
km s$^{-1}$. Fig.\,\ref{wings} shows that as we move away from the
line core, the $K_1$ velocity exponentially declines to 120$\pm$10 km
s$^{-1}$, while the systemic velocity, $\gamma$, moves between 100 and
123 km s$^{-1}$. The phasing for the compact object showed a small
change over Gaussian separation, but corresponds well with the value
predicted by the ephemeris derived above (i.e. $\simeq$0.5). Although
we expect this kind of behaviour for the $K_1$ velocity we are still
hesitant to identify this as the true motion of the compact object,
and we should treat the result with caution.

\subsection{Doppler maps}

Doppler tomography is a powerful tool for probing the structure of the
accretion disk (Marsh \& Horne 1988). It makes use of all the
available data and is effective for emission features that are too
weak to be identified in the individual spectra, by resolving the
distribution of the line emission in the co-rotating frame of the
binary system. The definition of orbital phase 0 in Doppler maps
corresponds to inferior conjunction of the secondary (Marsh 2001), and
is one of the main reasons why we changed the definition of our phase
0 compared to that adopted by Schaefer (1990). This should project a
signature of the donor star along the positive y-axis (Steeghs \& Casares
2002).

\begin{figure*}
\parbox{5.8cm}{\psfig{figure=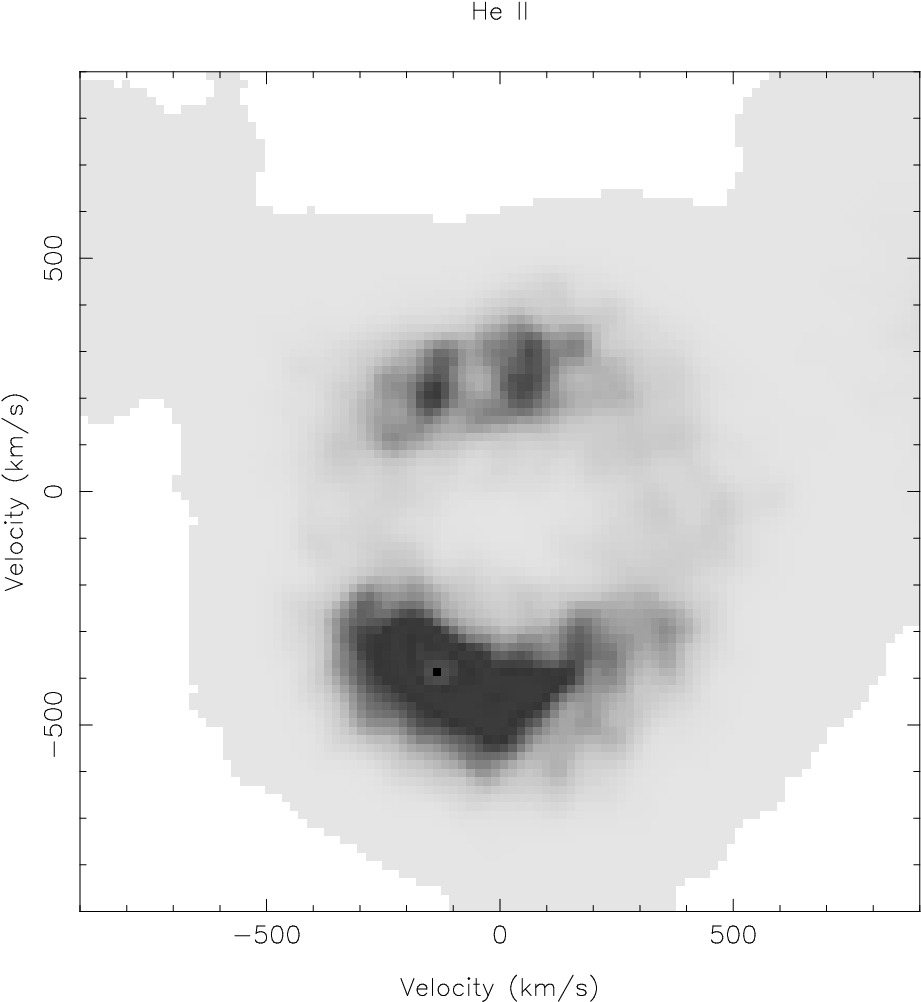,angle=-90,width=6.8cm}}
\parbox{5.8cm}{\psfig{figure=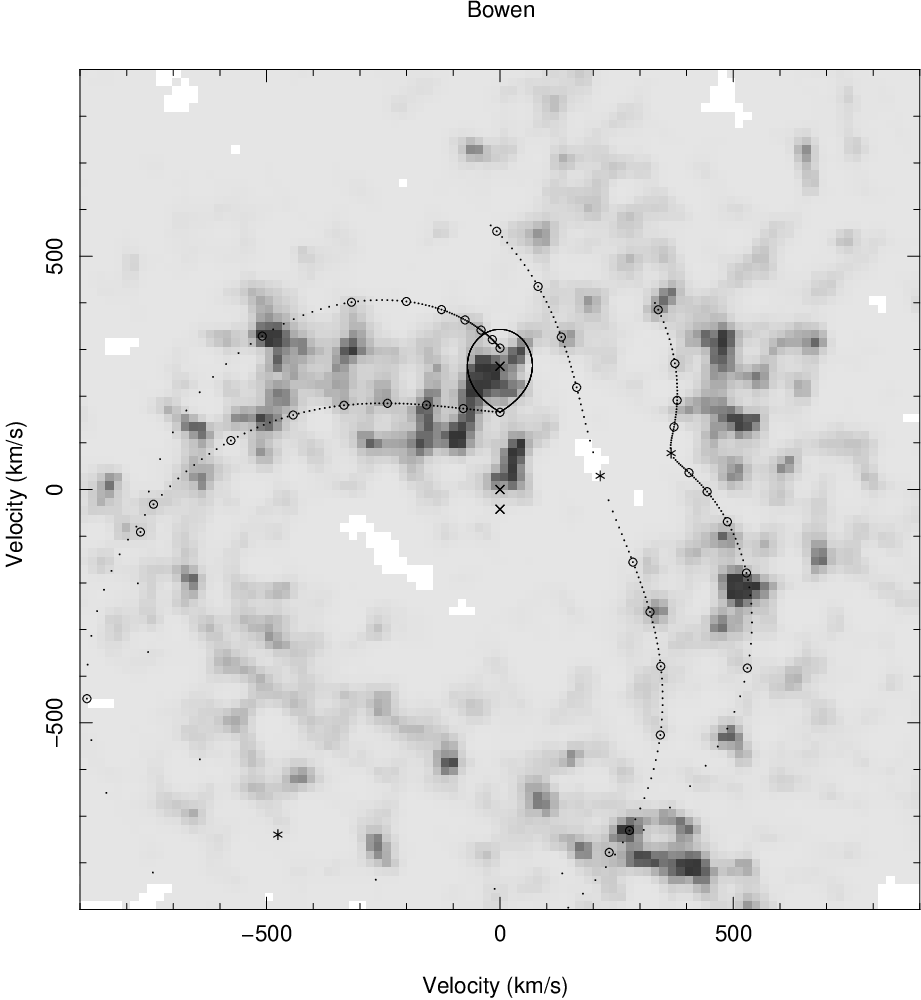,angle=-90,width=6.8cm}}
\vspace*{0.2cm}
\parbox{5.8cm}{\psfig{figure=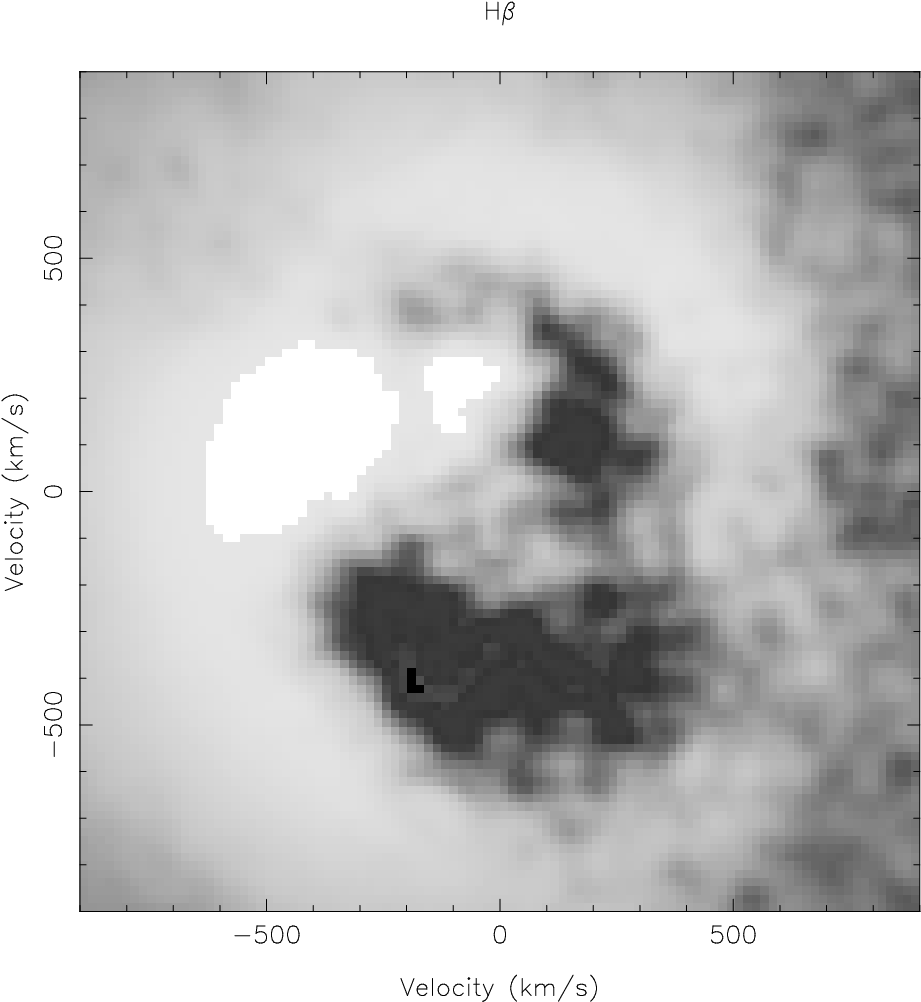,angle=-90,width=6.8cm}}
\parbox{5.8cm}{\psfig{figure=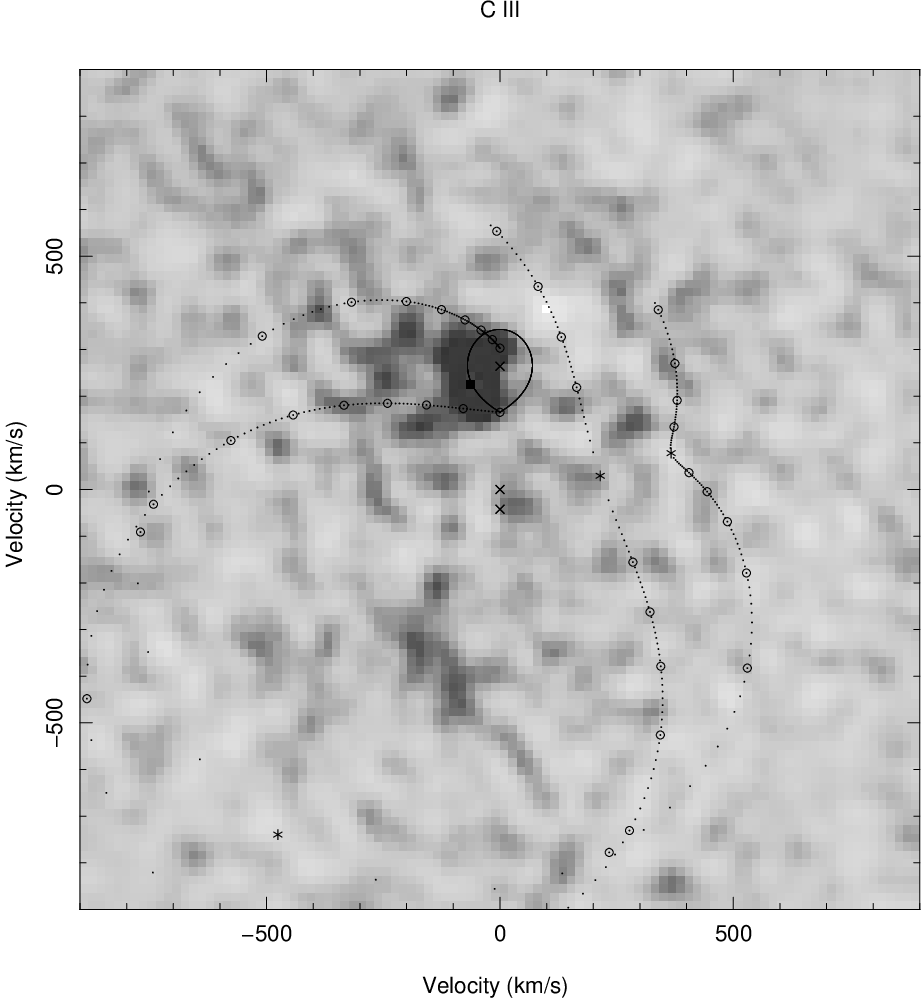,angle=-90,width=6.8cm}}
\caption{Doppler maps of He\,II $\lambda$4686 (top-left), the Bowen
region (top-right). The bottom panel shows the Doppler deconvolution of
the H$\beta$ (bottom-left) and C\,III $\lambda$4860 components as explained 
in Sect.\,3.5. In He\,II, Bowen and C\,III the white parts indicate
the zero level of the flux, while in H$\beta$ the white parts indicate
regions of most absorption. Indicated on the Bowen and C\,III maps
are the Roche lobe, gas stream leaving the L1 point, and the Keplerian
velocity along the stream for the set of parameters $q$=0.16 and 
$K_2$=264 km s$^{-1}$ discussed in Sect.\,4.2.
\label{doppler}}
\end{figure*}

Fig.\,\ref{doppler} shows the Doppler maps of the three most important
emission line regions, i.e. He\,II $\lambda$4686, the Bowen blend and
H$\beta$. Since there are no other emission lines present around
He\,II, we could use the standard reduction techniques with a systemic
velocity of 90 km s$^{-1}$. As expected from the trails (see
Sect.\,3.4) two main emission regions can be observed
(Fig.\,\ref{doppler}/top-left), more or less in anti-phase with each
other. The bright region in the top half corresponds with the position
where the donor star and accretion stream are expected to be located.
However, the brightest feature is the extended region in the
bottom-left quadrant of the Doppler map, a position where no enhanced
emission is expected for a 'normal' X-ray binary.

Using the correct systemic velocity should produce the sharpest
defined feature in the top part of the Doppler map. We therefore tried
different velocities with steps of 10 km s$^{-1}$ around 90 km
s$^{-1}$. However, in line with Sect.\,3.4, we found that 90 km
s$^{-1}$ gives the sharpest result for the compact spot in the top
part of the map, and again interpreted this as the
systemic velocity. Furthermore, we created Doppler maps of He\,II
$\lambda$4686 for each individual night in order to explore any
significant change. All maps looked identical, and no long-term change
in the emission line profile appears to be present.

The Doppler map of the Bowen region (Fig.\,\ref{doppler}/top-right)
was created using standard reduction techniques, but fitting the three
strongest C\,III lines (4647,4650 and 4665 \AA) simultaneously using
relative strengths as indicated by McClintock et~al. (1975). The most
pronounced feature in the map is a sharp spot in the top half of the
map, at a position similar to that in the He\,II map and where the
donor star is expected. Fitting a 2-dimensional Gaussian to this spot
we measure a velocity of 250$\pm$25 km s$^{-1}$. This 25 km s$^{-1}$
is the formal 1-sigma fitting error in the centroid position of our
2-dimensional Gaussian. Although the spot is offset from the x-axis by
35 km s$^{-1}$, this only corresponds to a phase offset of 0.02 which
is smaller than the error in the phase origin (0.04). The spot
location is thus consistent with being along the axis connecting the
neutron star with the donor star. We also rotated the map by 0.02
orbital phase in order to align the spot with the x-axis. Since we
measured a similar velocity as above, we concluded that the error
introduced by the off-set is negligible.

The most complex region is H$\beta$, where  C\,III
$\lambda$4860/He\,II $\lambda$4859 and an H$\beta$ emission line are
superposed on a deep absorption feature. Since it is currently not
possible to directly create Doppler maps with negative pixel values we
used a non-standard reduction technique. First we created a Doppler
image that consisted of a Gaussian with a peak and width comparable to
the observed absorption feature. From this image we created artificial
data that has the same properties as the observations, i.e.  creating
a trail of Gaussians. We then added the artificial data to the
observations in order to remove the absorption feature. We then used
standard analysis techniques to create Doppler images for both the
C\,III/He\,II and H$\beta$ simultaneously. Finally, we
subtracted the artificial map of the Gaussian to produce the maps
shown in Fig.\,\ref{doppler} (bottom).  Although we were able to more
or less disentangle the different components, we must be cautious
about the details of the two maps.

\begin{figure}
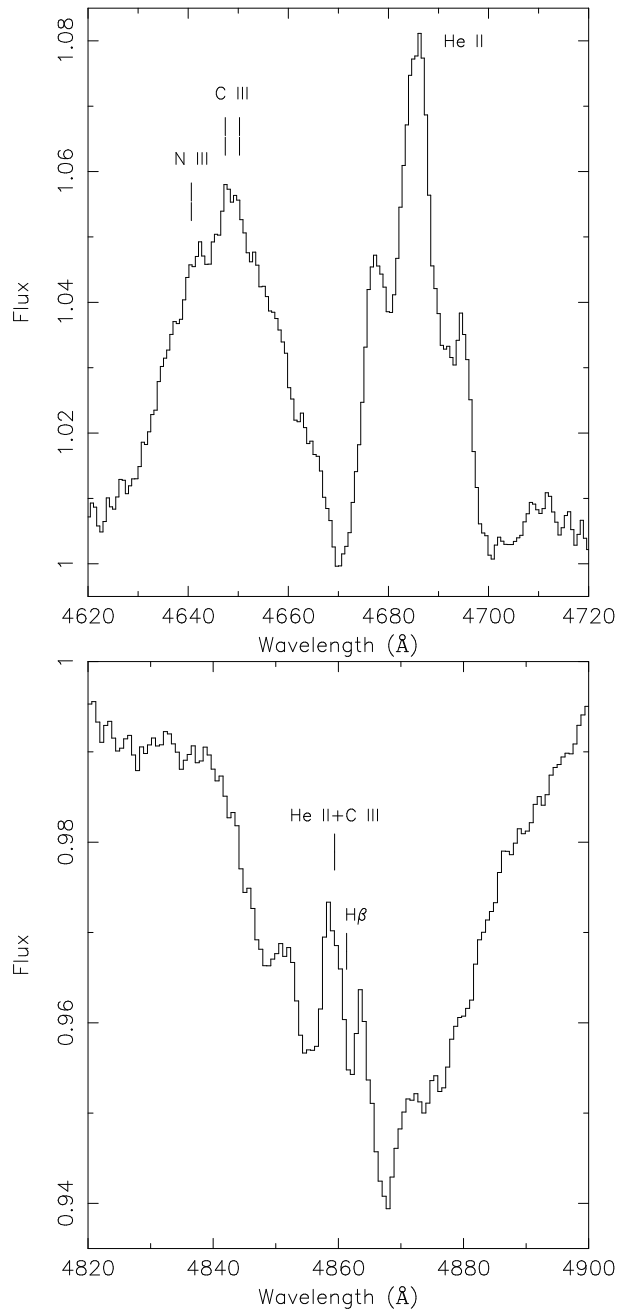
\begin{center}
\psfig{figure=gx9_bowen_rest.ps,angle=-90,width=8.0cm}
\psfig{figure=gx9_hbeta_rest.ps,angle=-90,width=8.0cm}
\caption{Average spectrum of the Bowen (top) and H$\beta$ (bottom)
regions of GX\,9$+$9 in the rest-frame of the donor star. We have
indicated the dominant contributing lines in these regions.
\label{rest}}
\end{center}\end{figure}

The C\,III/He\,II Doppler map (Fig.\,\ref{doppler}/bottom-right) looks
remarkably similar to that of the Bowen region, suggesting that the
line is indeed dominated by C\,III emission. The main feature is a
single localised spot in the top half of the map, that is at a
velocity of 210$\pm$25 km s$^{-1}$ (and an off-set from the x-axis of
$\simeq$55 km s$^{-1}$, still within our phase zero uncertainty),
consistent with the spot in the Bowen map. The H$\beta$ map
(Fig.\,\ref{doppler}/bottom-left) looks more complex. All absorption
is attributed to H$\beta$ and is present at all orbital phases,
although it is most prominent in the left part of the map around
$y$=0, $x$=-500 km s$^{-1}$. Note that in this map the lightest
regions indicate negative flux levels, in sharp contrast with the
other maps where lightest regions represent the continuum level.
However, the most prominent feature in the H$\beta$ map is created by
the emission lines, namely the extended region in the bottom half of
the map that has a comparable shape and position to that in the He\,II
map. However, this similarity between the H$\beta$ and He\,II maps is
rather curious, since in other LMXBs these lines can show rather
different characteristics (e.g. Steeghs \& Casares 2002). Although
this might be explained by contamination of He\,II $\lambda$4859, we
would have expected that this would mainly affect the C\,III map. We
must, therefore, conclude that this emission region in the bottom half
of the H$\beta$ map is likely real.

\section{Discussion}

We have presented medium resolution optical spectroscopy of GX\,9$+$9.
One of the most remarkable characteristics of GX\,9$+$9 is the strong
C\,III emission in its spectrum, while N\,III is hardly detected. To
our knowledge, no other X-ray binary is known that shows such an
unusual C\,III/N\,III ratio in the Bowen region.  For example, in
X1822$-$371 and Sco\,X-1 the narrow Carbon and Nitrogen lines in the
Bowen region were more or less of comparable strength, and no other
strong C\,III/N\,III lines could be detected (Casares et~al. 2003,
Steeghs \& Casares 2002). This could give us important constraints on
the evolutionary status of GX\,9$+$9. However, we can exclude a more
evolved core in the donor star of GX\,9$+$9, since this would lead to
an overabundance of N\,III due to processing by the CNO cycle as has
been observed in e.g. XTE\,J1118$+$480 Haswell et~al. (2002). The only
other systems that show prominent Carbon emission lines in their
spectra are some of the ultra-compact X-ray binaries (Nelemans et~al.
2006). However, in these systems the donor star is thought to be
a white dwarf that is deficient of both H and He.  These systems are
clearly different to GX\,9$+$9, which is not an ultra-compact X-ray
binary and has both H and He in its spectrum.

\begin{table}\begin{center}
\caption{Derived system parameters for GX\,9$+$9.
\label{system}}
\begin{tabular}{lclc}
\hline 
Parameter & & Parameter &\\ 
\hline 
$P_{\rm orb}$ (hrs)        & 4.1934528(20)    & $f(M)$ ($M_\odot$)  & ($\ge$)0.22$\pm$0.03\\ 
$T_0$ (HJD)                & 2 453 151.464(3) & $q$ ($M_2$/$M_1$)   & 0.07-0.35\\
$\gamma$ (km s$^{-1}$)     & 90$\pm$7         & $K_2$ (km s$^{-1}$) & 182-406\\
$K_{\rm em}$ (km s$^{-1}$) & 230$\pm$35       &\\
\hline
\end{tabular}
\end{center}\end{table}

\subsection{Orbital Period and Long Term Variations}

Following Levine et~al. (2006a,b) we have updated the ephemeris of
GX\,9$+$9 using the RXTE/ASM data. Given the coincidence of optical
minimum with the minimum in the X-ray lightcurve, as well as the
presence of Bowen emission spots on the y-axis consistent with the
donor star, we feel confident that this ephemeris tracks the correct
orbital phase of the binary. Levine et~al. (2006a) showed that the
strength of the modulation in GX\,9$+$9 increased from 6\% before Jan
2005 to 18\% after this date (and until at least June 2006). A
possible explanation for this change in modulation is given by the
folded RXTE/ASM lightcurve shown in Fig.\,\ref{folded}. There appears
to be a partial eclipse around orbital phase 0, i.e. at inferior
conjunction of the secondary. The most likely region that could be
eclipsed by the secondary is the extended accretion disk corona that
Kong et~al. (2006) observed in GX\,9$+$9, suggesting that the corona
slowly changes size. One option could be that the orbital modulation
in X-rays disappears when the corona becomes too small to be eclipsed
by the secondary. An alternative option is that the corona is always
eclipsed, but when it increases in size its fractional contribution to
the total X-ray flux increases thereby lowering the depth of the
eclipse.  If the size of this corona could be determined, we would be
able to distinguish between the different scenarios, and could also
constrain the inclination of GX\,9$+$9. However, the fact that this
partial eclipse occurs suggests a reasonably high inclination.

Haswell et~al. (2001) proposed that GX\,9$+$9 is a persistent
superhumper. Superhumps are periodic optical modulations first
observed during outburst of SU\,UMa dwarf novae (see Warner 1995 for a
review), and their most important property is that the photometric
period is a few percent (1-7\%) longer than the orbital period.  This
is thought to be due to prograde precession of an eccentric accretion
disk, giving rise to slightly longer periods (with respect to the
orbital period) for the accretion disk. The main criterion
for creating an eccentric disk is an extreme mass ratio
($M_2$/$M_1$$<$0.33). Comparing our X-ray period with the photometric
period derived by Kong et~al. (2006) we note that the photometric one
is 0.0564 hrs longer, corresponding to only $\simeq$1\% of the orbital
period. Although this difference is small, it is still significant
according to the formal errors (4.7$\sigma$), and we therefore
further investigated the possibility of a superhump in GX\,9$+$9.

\begin{figure}\begin{center}
\psfig{figure=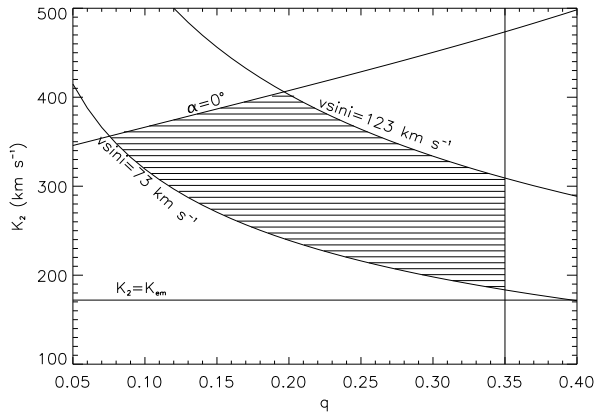,width=8.0cm}
\caption{Constraints on $q$ and $K_2$ for GX\,9$+$9. $K_2$ must be
larger than $K_{\rm em}$, and a disc opening angle of $\alpha$=0 gives
the upper-limit for $K_2$. The rotational broadening derived from the
emission line widths sets a lower limit to $q$, while the largest
possible main sequence star that fits the Roche-lobe of GX\,9$+$9,
combined with its maximum allowed broadening sets the upper-limit to
$q$. All curves indicate 95\% confidence levels.
\label{limits}}
\end{center}\end{figure}

If GX\,9$+$9 is a superhumper, we would also expect similar
spectroscopic characteristics for GX\,9$+$9 to that of the known
superhumping LMXB XTE\,J1118$+$480, namely the movement of the nightly
mean emission line profiles (e.g. Zurita et~al. 2002). Such movement
is not observed in GX\,9$+$9, arguing against a superhump
interpretation.  Furthermore, also the so-called $\epsilon$($q$)-$q$
relation, where $q$ is the mass ratio and $\epsilon$($q$) the
fractional period excess of the superhump over the orbital period,
suggest that the photometric period cannot be due to a superhump.  If the
difference in photometric and X-ray period is real, the empirical
relation by Patterson et~al. (2005) suggests that $q$$\simeq$0.003, an
unrealistically small value (as we will show in Sect.\,4.2). This
could indicate that the errors on the orbital periods have been
underestimated, and that the X-ray and photometric period are
similar. On the other hand we cannot completely exclude the
possibility that the movement of the nightly mean emission line
profile is hidden by the strong emission feature in the lower-left
quadrant of the Doppler maps (Fig.\,\ref{doppler}), and that the
$\epsilon$($q$)-$q$ does not apply to persistently bright LMXBs. The
only thing we therefore conclude is that there are no indications of a
superhump in our dataset, and that a better determination of the
photometric period is needed to find out if the difference with the
X-ray period is real.

\subsection{Donor Star and System Parameters of GX\,9$+$9}

Thus far narrow lines in the Bowen region have been detected in
Sco\,X-1, X\,1822$-$371, GX\,339$-$4, 4U\,1636$-53$, 4U\,1735$-$44 and
Aql\,X-1 (Steeghs \& Casares 2002; Casares et~al. 2003, Hynes et~al.
2003, Casares et~al. 2006, Cornelisse et~al. 2007). We can now also
add GX\,9$+$9 to this list. Phase zero of the radial velocity curve of
these narrow lines in GX\,9$+$9 correspond well with the minima in the
X-ray and optical lightcurves, suggesting that these lines arise close
to or are connected with the donor star. It has been proposed for the
other sources that these narrow lines arise on the surface of the
secondary, and especially for the eclipsing LMXB X\,1822$-$371 this
connection could unambiguously be claimed (Casares. 2003). Since it is
possible, for a valid set of system parameters (see below), to draw a
Roche lobe on the Bowen and C\,III Doppler maps that encompasses most
of the observed narrow spots (see Fig.\,\ref{doppler}), we propose
that also in the case of GX\,9$+$9 these lines arise on the donor
star. If our suggestion is correct, the narrow feature in the Doppler
maps of both the Bowen region and C\,III $\lambda$4860 (see
Fig.\,\ref{doppler}) must be produced somewhere on the inner
hemisphere of the companion, and has a $K_{\rm em}$ of 230$\pm$35 km
s$^{-1}$ (the average value derived from the compact spots in the
Doppler maps of He\,II $\lambda$4686, Bowen and C\,III $\lambda$4860).
However, given that only the irradiated side facing the compact object
is contributing to this emission, this should be considered a
lower-limit to the velocity semi-amplitude of the centre of mass of
the donor $K_2$.

\begin{figure}
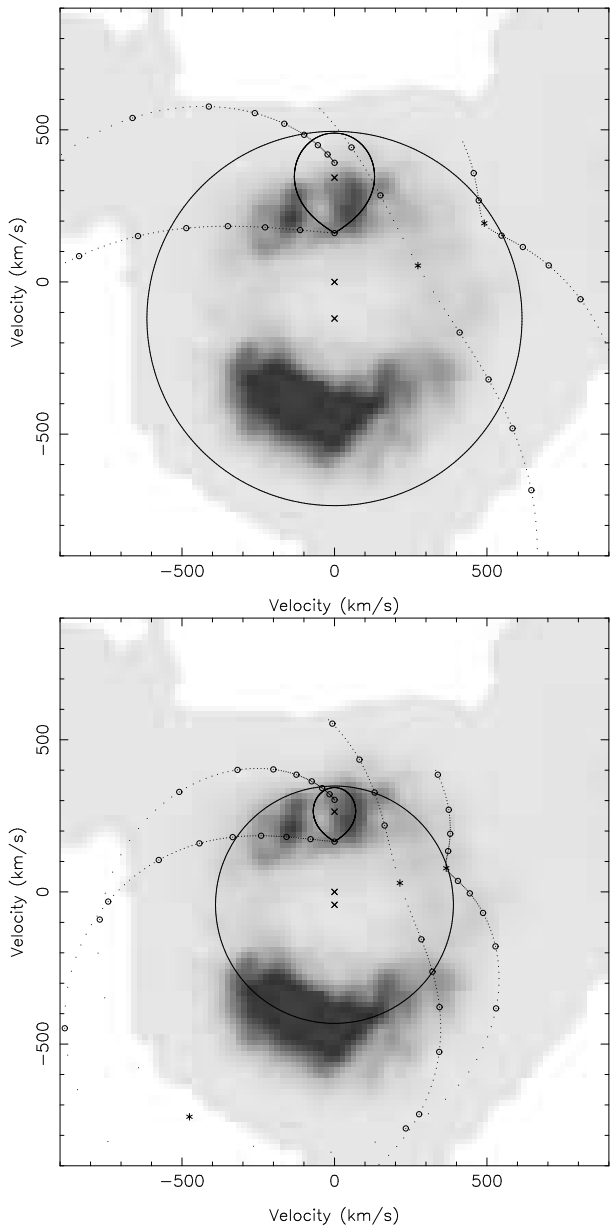
\begin{center}
\psfig{figure=he_stream1.ps,angle=-90,width=8.0cm}
\psfig{figure=he_stream2.ps,angle=-90,width=8.0cm}
\caption{He\,II $\lambda$4686 Doppler map showing the Roche lobe of
the secondary, the gas stream leaving the L1 point, and the Keplerian
velocity along the stream for two different sets of system parameters.
The circle indicates the Keplerian velocity at the edge of the disk. 
Top: System parameters derived for the assumption that $K_1$ derived with
the double Gaussian technique is correct.
Bottom: System parameters derived using the assumption that the emission 
feature in the bottom left quadrant is part of the accretion disk.
\label{stream}}
\end{center}\end{figure}

In order to derive some constraints on the system parameters of
GX\,9$+$9, we can start by determining a lower limit on the mass
function of $f(M)$=$M_1\sin^3i/(1+q)^2$$>$0.22$\pm$0.03$M_\odot$
(where $q$=$M_2$/$M_1$ is the mass ratio and $i$ the inclination of
GX\,9$+$9). This lower limit corresponds with the assumption that
$K_{\rm em}$=$K_2$, i.e. that the narrow lines are produced on the
poles of the donor star. In order to find the real $K_2$ we need to
determine the $K$-correction, which is dependent on the disk flaring
angle ($\alpha$), $q$, and weakly on $i$ (see Mu\~noz-Darias et~al.
2005). Using an irradiation binary code, Mu\~noz-Darias et~al.
determined 4th order polynomials for different values of $\alpha$ and
$i$ to estimate the $K$-correction. Unfortunately we do not have any
solid estimates of $K_1$, since the double Gaussian method used in
Sect.\,3.4 is not valid for an accretion disk that is dominated by a
hotspot. Therefore we cannot directly apply the $K$-correction, and
must therefore resort to several assumption in order to further
constrain the system parameters. Although we will point out what the
major source of uncertainty is for each of our assumptions, we do
think that they are currently the best we can do.

First of all, the simplest assumption is that $K_2$$\ge$$K_{\rm em}$,
and show the 95\% confidence limit in Fig.\,\ref{limits}.
Furthermore, we can use the polynomials by Mu\~noz-Darias et~al.
(2005) for $\alpha$=0$^{\circ}$ (and $i$=40$^\circ$) to determine the
upper-limit to $K_2$ as a function of $q$. Again we show the 95\%
confidence curve in Fig.\,\ref{limits}.

We can determine a lower-limit to $q$ by estimating a lower limit to
the rotational broadening, $V$$\sin i$, of the emission lines produced
on the surface of the donor (Wade \& Horne 1988). However, this does
mean that we assume that all of the line broadening is due to
rotation, which might not be the case. We created an average spectrum
in the rest-frame of the secondary. Fig.\,\ref{rest} shows the two
most prominent regions (around He\,II $\lambda$4686 and H$\beta$) of
the rest-frame spectrum. Unfortunately, most lines are either too
weak, are located in very complicated regions (e.g. Bowen and H$\beta$
regions) or have a component that does not originate on the secondary
(e.g. He\,II $\lambda4686$). We decided to estimate the {\it FWHM} of
the two C\,III lines in the Bowen region, C\,III $\lambda$4859, and
also the H$\beta$ absorption line. All lines give a similar {\it FWHM}
that is on average 153$\pm$26 km s$^{-1}$.  However, we do note that
this includes the effect of the intrinsic broadening due to the
instrumental resolution of 70 km s$^{-1}$. To take this into account
we broadened a strong line in our arc spectrum using a Gray rotational
profile (Gray 1992) until we reached the observed {\it FWHM}. We
assumed no limb-darkening since the fluorescence lines occur in
optically thin conditions. We found that a rotational broadening of
102$\pm$15 km s$^{-1}$ reproduces our results. Using $V_{\rm
  rot}$$\sin$$i$=0.462$K_2$$q^{1/3}$$(1+q)^{2/3}$ (Wade \& Horne
1988), and $V$$\sin$$i$$\ge$73 km s$^{-1}$ (95\% confidence) we can
draw this curve in Fig.\,\ref{limits}.

We also show the limit corresponding to the assumption that the donor
star is a zero-age main sequence star that fits in a 4.2 hrs period
Roche lobe. Although there is no guarantee that the donor star in
GX\,9+9 satisfies that assumption (see e.g. Schenker \& King 2002 for
alternative scenarios), we include it in Fig.\,\ref{limits} as a
likely constraint on the system parameters (Casares et~al. 2006).
This leads to a donor star mass $\le$0.47$M_\odot$ and together with
$M_{\rm NS}$$\ge$1.35$M_\odot$ gives $q$$\le$0.35. Note that this
implies that the lower limit on $K_2$ derived from the compact spot in
the Bowen and C\,III Doppler maps is always below the rotational
broadening curve determined from the widths of the emission lines.

By estimating the upper-limit to the rotational broadening for a
0.47$M_\odot$ main sequence star donor we derive the final limit shown
in Fig.\,\ref{limits}. From Paczynski (1974, 1983) we can determine an
upper-limit to the inclination of $\le$65$^\circ$, i.e. where the
donor star would just eclipse the compact object.  Since Kong et~al.
(2006) did not detect any orbital variation in X-rays, and their X-ray
spectral modelling showed that the emission coming from the inner
accretion disk or boundary layer is a significant fraction of the
total flux ($\simeq$25\%), we think our assumption on the inclination
is valid. This gives a maximum rotational broadening of
$V$$\sin$$i$$=$123 km s$^{-1}$, as indicated in Fig.\,\ref{limits}.

The values for the system parameters of GX\,9$+$9 we have derived are
the best limits possible, and these are shown in Fig.\,\ref{limits}
and tabulated in Table\,\ref{system}. In the remainder of this section
we will speculate in two different ways about further constraints on
the system parameters using more uncertain assumptions. Although they
lead to very different system parameters, we will use the results of
these speculations for our discussion of the accretion disk structure
in Sect.\,4.3.

Although the double Gaussian technique is not an appropriate way to
determine $K_1$ in the case of GX\,9$+$9, the value derived is within
the acceptable limits. We can therefore speculate that this is the
correct $K_1$ and see how this limits the system parameters. Again
using the polynomials by Mu\~noz-Darias (2006) we can determine the
largest $K$-correction, which leads to $K_2$$\le$369 km s$^{-1}$ and
thereby $q$$\ge$0.32. Since we already established that $q$ must be
smaller than 0.35 (see Fig.\,\ref{limits}), we derive a lower limit to
$K_2$ of 343 km s$^{-1}$ in this case. We note that these system
parameters are all toward the upper-limits of the allowed area in
Fig.\,\ref{limits}. We use the lower limits of our derived parameters
($q$=0.35 and $K_2$=343 km s$^{-1}$) to draw the Roche-lobe of the
secondary, stream trajectories and outer edge of the accretion disk on
the Doppler map of He\,II and show this in Fig.\,\ref{stream} (top
panel). We have indicated the Roche lobe of the secondary and the
outer edge of the accretion disk as solid lines. The crosses indicate
the velocities of the centre of mass, and the two stars, while the
dotted lines indicate the ballistic trajectory of the material leaving
the L1 point and the Keplerian velocity the ballistic stream passes
through. It is clear from this figure that in this case the emission
feature in the lower-left quadrant is at too low velocities to be
associated with the accretion disk.

Another way to estimate the system parameters is if we speculate that
the emission feature in the lower-left quadrant must be part of the
accretion disk. Furthermore, we can also assume that the feature close
to the donor star signature is part of the accretion stream. These
constraints imply a small $K_2$ (to make the velocities of the outer
edge of the disk similar to the feature in the lower-left) and also a
small $q$ (the Roche-lobe of the secondary must be small to also let
the stream pass through the feature close to the donor star). Assuming
that our estimate of $V$$\sin$$i$ is correct, there is a possible
combination of $q$ and $K_2$ at the lower limits of our allowed region
($q$=0.16 and $K_2$=264 km s$^{-1}$ for $\alpha$=12$^\circ$ and
$i$=40$^\circ$) that fulfils all our criteria. and we show the result
in Fig.\,\ref{stream} (bottom).
 
Note that we do not claim that either of the two sets of system
parameters shown in Fig.\,\ref{stream} must be correct. Given the
large area of possible solutions in Fig.\,\ref{limits} it would be
very unlikely that they are, but they nicely illustrate that using
different assumptions, both not too unreasonable, it is possible to
derive completely different system parameters. In order to really
constrain the system parameters of GX\,9$+$9, it is necessary to
unambiguously determine $K_1$, as was for example possible for
X\,1822$-$371 or V801\,Ara (Jonker \& van der Klis 2001; Casares
et~al. 2006).  Unfortunately, this is going to be difficult for
GX\,9$+$9, since thus far it has not been possible to directly observe
a signal coming from the compact object (in the form of pulsations or
Type\,I X-ray bursts) or even the most inner regions of the accretion
disk (in the form of high frequency quasi-periodic oscillations (e.g.
van der Klis 2006).

\subsection{Accretion Disk Structure}

GX\,9$+$9 shows several interesting characteristics that are not
commonly observed in LMXBs. Most important of all is the presence of a
strong emission feature in the lower-left quadrant of both the He\,II
$\lambda$4686 and H$\beta$ Doppler maps (see Fig.\,\ref{doppler}),
  while another uncommon feature is the extremely broad absorption
  feature around H$\beta$ that is present at all phases.

In many LMXBs, the He\,II Doppler map displays a classic disk-like
structure, with sometimes a spot in the upper-left quadrant due to the
accretion stream impacting the disk. However, having emission
predominantly in the lower-left quadrant of the Doppler map is not
unique for GX\,9$+$9.  There are at least two LMXBs that show a
similar He\,II Doppler map, the transient sources XTE\,J2123$-$058S
(Hynes et~al. 2001) and GRO\,J0422$+$32 (Casares et~al. 1995). In
particular, a comparison between the He\,II $\lambda$4686 trails of
GX\,9$+$9 and XTE\,J2123$-$058 shows that they are strikingly similar
to each other. Hynes et~al. (2001) compared the properties of
XTE\,J2123$-$058 with that of the SW\,Sex objects, a subclass of
nova-like cataclysmic variables (see e.g. Thorstensen et~al 1991).
Although there are several competing models, it is generally thought
that SW\,Sex objects have some sort of stream (that is not connected
to the accretion disk), that is either overflowing the disk or being
propelled away, that gives rise to their peculiar characteristics (see
e.g. Hoard et~al. 2003, Hellier 2000, Horne 1999).

In Fig.\,\ref{stream} we show various system characteristics
superposed on the He\,II Doppler map for two different sets of system
parameters. The important thing to notice in Fig.\,\ref{stream}
(bottom) is that there are combinations of valid system parameters
possible where the emission feature in the lower-left quadrant
overlaps with the expected velocities for the outer edge of the
accretion (although some emission it is still at sub-Keplerian
velocities). This suggests that most of the He\,II emission is
produced by material that is carried downstream after the stream
impacts the accretion disk. Such behaviour was also observed in
EXO\,0748$-$676 where especially emission from the higher ionization
emission lines showed up further downstream (Pearson et~al. 2006), at
a position comparable to the He\,II emission in GX\,9$+$9. However,
there are also many combinations of the system parameters possible
where all He\,II emission in GX\,9$+$9 is at sub-Keplerian velocities
(as illustrated in Fig.\,\ref{stream}), and that disk overflow must
take place.  In this case it would be very similar to
XTE\,J2123$-$058, and it would then be the second LMXB to show SW\,Sex
behaviour (Hynes et~al. 2001). Unfortunately, currently we cannot
distinguish between both scenarios, and must conclude that both are
just as likely.

The presence of the broad absorption component around H$\beta$ in
GX\,9$+$9 is another curious feature. Although other LMXBs also show
such a feature, e.g. X\,1822$-$371, Ser\,X-1 and XTE\,J2123$-$058
amongst others (Casares et~al.  2003, Hynes et~al. 2004, Hynes et~al.
2001), it is usually not as strong as observed in GX\,9$+$9.  An
exception is the X-ray transient GRO\,J0422$+$32 (that showed a
similar He\,II $\lambda$4686 Doppler map as GX\,9$+$9) where the
absorption feature around H$\beta$ (and H$\gamma$) shows a similar
strength (Casares et~al. 1995). Although this might be a coincidence,
since XTE\,J2123$-$058 does not show such prominent absorption (Hynes
et~al.  2001), it is curious. An explanation could be a high
inclination for GX\,9$+$9 that causes absorption by the optically
thick parts of the inner accretion disk. This is supported by
the shape of the folded X-ray lightcurve (Fig.\,\ref{folded}) and the
fact that Kong et~al. (2006) concluded that a significant fraction of
the inner accretion disk or boundary layer is not visible due to
obscuration by the inner or thickened disk.  Interestingly,
XTE\,J2123$-$058 and many SW\,Sex objects also have a high inclination
(Casares et~al. 1998, Knigge et~al. 2000). Although a high inclination
has also been suggested for GRO\,J0422$+$32 (Kato et~al. 1995), there
are indications that it is lower (e.g. Gelino \& Harrison.  2003;
Reynolds et~al. 2007).

\section{Conclusions}

We have presented a spectroscopic dataset of GX\,9$+$9, and using the
Bowen fluorescence technique have for the first time detected narrow
lines that we interpret as coming from the donor star. If true, this
allowed us to determine a lower limit on the mass function of the
system of 0.22$M_\odot$ and 182$\le$$K_2$$\le$406 km s$^{-1}$. In
order to constrain the system parameters more information is
necessary, in particular on $K_1$.  However, given that thus far no
unambiguous detection of the compact object has been made this is
going to be difficult.

We have updated the ephemeris of GX\,9$+$9 using RXTE/ASM X-ray data,
and the folded lightcurve suggests a partial eclipse of the extended
accretion disk corona by the secondary. The Doppler maps of GX\,9$+$9
show the presence of stream overflow.  This is either of material that
is carried downward along the rim of the accretion disk, or
overflowing material that is at sub-Keplerian velocities comparable to
XTE\,J2123$-$058 and SW\,Sex systems (Hynes et~al. 2001). Finally,
GX\,9$+$9 shows a curious C\,III/N\,III ratio compared to other LMXBs,
which could provide an important hint about the evolutionary status of
this system.

\section*{Acknowledgements}
Based on data collected at the European Southern Observatory Paranal,
Chile (Obs.Id. 073.D-0819(A)). We would like to thank the referee,
Peter Jonker, for the careful and helpful comments wich have improved
this paper. We would like to thank the RXTE/ASM teams at MIT and GSFC
for provision of the on-line ASM data. We acknowledge the use of
PAMELA, MOLLY and DOPPLER developed by T.R.  Marsh. We acknowledge the
use of the on-line atomic line list at
http://www.pa.uky.edu/$\sim$peter/atomic. RC acknowledges financial
support from a European Union Marie Curie Intra-European Fellowship
(MEIF-CT.2005-024685). JC acknowledges support from the Spanish
Ministry of Science and Technology through project AYA2002-03570. DS
acknowledges a Smithsonian Astrophysical Observatory Clay fellowship
as well as support from NASA through its guest Observer program. DS
acknowledges a PPARC/STFC Advanced Fellowship.

\bsp

\label{lastpage}

\end{document}